\documentclass[a4paper,11pt,]{article}

\usepackage[left=2.5cm,
            right=2.5cm,
            top=2.5cm,
            bottom=2.5cm]{geometry}

\usepackage[T1]{fontenc}
\usepackage[utf8]{inputenc}
\usepackage[english]{babel}
\usepackage{amsmath,amsfonts,amssymb}
\usepackage{lmodern}
\usepackage[scaled]{helvet}
\usepackage[T1]{fontenc}
\usepackage[scaled=0.95]{inconsolata}

\usepackage[dvipsnames]{xcolor}
\definecolor{black50}{gray}{0.5} 
\definecolor{color0}{RGB}{0,0,0} 
\definecolor{color1}{RGB}{59,90,198} 
\definecolor{pnasbluetext}{RGB}{0,101,165}
\definecolor{pnasblueback}{RGB}{205,217,235}

\definecolor{macblue}{HTML}{185FAF}  
\definecolor{pnasbluetext}{RGB}{0,0,0} %

\usepackage{afterpage}

\usepackage{fontawesome5}  

\usepackage{afterpage}
\usepackage{ifpdf,ifxetex}
\ifpdf\else
  \ifxetex\else
    
    \pdfpagewidth=\paperwidth
    \pdfpageheight=\paperheight
\fi\fi
\usepackage{tikz}
\usepackage[framemethod=tikz]{mdframed}

\IfFileExists{flushend.sty}
    {\usepackage{flushend}}
    {\typeout{}
    \typeout{Package widetext error: Install the flushend package which is
    a part of sttools bundle. Available from CTAN.}
    \typeout{}
    \stop
    }

\IfFileExists{cuted.sty}
    {\usepackage{cuted}}
    {\typeout{}
    \typeout{Package widetext error: Install the cuted package which is
    a part of sttools bundle. Available from CTAN.}
    \typeout{}
    \stop
    }

{%
    \par
    \hfill\rule[-6pt]{0.4pt}{6.4pt}%
    \rule{\dimexpr(0.5\textwidth-0.5\columnsep-1pt)}{0.4pt}
    \end{strip}
}

\usepackage[colorlinks=true, allcolors=pnasbluetext]{hyperref}

\newcommand{\footerfont}{\normalfont\sffamily\footnotesize}
\newcommand{\titlefont}{\fontfamily{lmss}\bfseries\LARGE}

\newcommand{\addinfofont}{\normalfont\sffamily\fontsize{7}{8}\selectfont}
\newcommand{\absfont}{\normalfont\sffamily\small}
\newcommand{\keywordsfont}{\normalfont\sffamily\small}

\usepackage{xifthen}
\newboolean{shortarticle}
\newboolean{singlecolumn}

\def\useignorespacesandallpars#1\ignorespaces\fi{%
  #1\fi\ignorespacesandallpars}
\def\ignorespacesandallpars{%
  \@ifnextchar\par
  {\expandafter\ignorespacesandallpars\@gobble}%
  {}%
}

\newboolean{displaycopyright}
\setboolean{displaycopyright}{false} 
\usepackage{textcomp} 

\usepackage{graphicx,xcolor}
\usepackage{colortbl}
\usepackage{booktabs}
\usepackage[noend]{algpseudocode}
\usepackage{changepage}

\usepackage[labelfont={bf,sf}, labelsep=period]{caption}

\usepackage{natbib}
\DeclareCaptionFormat{pnasformat}{\normalfont\sffamily\small #1#2#3}
\usepackage{etoolbox}
\AtBeginEnvironment{tabular}{\small}

\usepackage{algorithm}
\usepackage[noend]{algpseudocode}
\algnewcommand\algorithmicinput{\textbf{Input:}}
\algnewcommand\algorithmicoutput{\textbf{Output:}}
\algnewcommand\Input{\item[\algorithmicinput]}%
\algnewcommand\Output{\item[\algorithmicoutput]}%
\algnewcommand\Step{\item[]}%

\usepackage{fancyhdr}  
\usepackage{lastpage}  
\def\footercontents#1{\def\@footercontents{#1}}
\footercontents{Package Vignette}
\newcommand{\printfootercontents}{\@footercontents}

\makeatletter
\fancypagestyle{firststyle}{
   \fancyfoot[R]{\footerfont \printfootercontents\hspace{7pt}|\hspace{7pt}\textbf{\@ifundefined{@documentdate}{\today}{\@documentdate}}\hspace{7pt}|\hspace{7pt}\textbf{\thepage\textendash\pageref{LastPage}}}
   \fancyfoot[L]{\footerfont\@ifundefined{@doifooter}{}{\@doifooter}}
}
\makeatother

\makeatletter
\appto{\titlefont}{\color{macblue}}

\usepackage[explicit,nobottomtitles]{titlesec}
\titleformat{\section}
  {\large\sffamily\bfseries\color{macblue}}
  {\thesection.}
  {0.5em}
  {#1}
  []
\titleformat{name=\section,numberless}
  {\large\sffamily\bfseries\color{macblue}}
  {}
  {0em}
  {#1}
  []
\titleformat{\subsection}
  {\sffamily\bfseries\color{macblue}} 
  {\thesubsection.}
  {0.5em}
  {#1}
  []

\titleformat{\subsubsection}[runin]
  {\sffamily\bfseries\color{macblue}}
  {\thesubsubsection.}
  {0.5em}
  {#1 \quad }
  []

\titleformat{\paragraph}[runin]
  {\sffamily\bfseries}
  {}
  {0em}
  {#1 }

\titlespacing*{\section}{0pc}{3ex \@plus4pt \@minus3pt}{10pt}
\titlespacing*{\subsection}{0pc}{2.5ex \@plus3pt \@minus2pt}{5pt}
\titlespacing*{\subsubsection}{0pc}{2ex \@plus2.5pt \@minus1.5pt}{2pt}
\titlespacing*{\paragraph}{0pc}{1.5ex \@plus2pt \@minus1pt}{12pt}

\newcommand{\additionalelement}[1]{\def\@additionalelement{#1}}
\newcommand{\addinfo}[1]{\def\@addinfo{#1}}
\newcommand{\doifooter}[1]{\def\@doifooter{#1}}
\newcommand{\documentdate}[1]{\def\@documentdate{#1}}
\newcommand{\leadauthor}[1]{\def\@leadauthor{#1}}
\newcommand{\etal}[1]{\def\@etal{#1}}
\newcommand{\keywords}[1]{\def\@keywords{#1}}
\newcommand{\authorcontributions}[1]{\def\@authorcontributions{#1}}
\newcommand{\authordeclaration}[1]{\def\@authordeclaration{#1}}
\newcommand{\equalauthors}[1]{\def\@equalauthors{#1}}
\newcommand{\correspondingauthor}[1]{\def\@correspondingauthor{#1}}
\newcommand{\significancestatement}[1]{\def\@significancestatement{#1}}
\newcommand{\matmethods}[1]{\def\@matmethods{#1}}
\newcommand{\acknow}[1]{\def\@acknow{#1}}

\def\xabstract{abstract}
\long\def\abstract#1\end#2{\def\two{#2}\ifx\two\xabstract
\long\gdef\theabstract{\ignorespaces#1}
\def\go{\end{abstract}}\else
\typeout{^^J^^J PLEASE DO NOT USE ANY \string\begin\space \string\end^^J
COMMANDS WITHIN ABSTRACT^^J^^J}#1\end{#2}
\gdef\theabstract{\vskip12pt BADLY FORMED ABSTRACT: PLEASE DO
NOT USE {\tt\string\begin...\string\end} COMMANDS WITHIN
THE ABSTRACT\vskip12pt}\let\go\relax\fi
\go}

\newcommand{\abscontent}{
\noindent
\parbox{\dimexpr\linewidth}{%
  \vskip3pt\absfont%
  {\bfseries Abstract} \theabstract
}%
\vskip10pt%
\noindent
\parbox{\dimexpr\linewidth}{%
{
 \keywordsfont
 \@ifundefined{@keywords}{}{{\bfseries Keywords} \@keywords}}%
}
\vskip12pt%
}

\newcommand{\abscontentformatted}{
\abscontent
}

\renewcommand{\@maketitle}{%
\def\And{\vskip5pt}
{%
\ifthenelse{\boolean{shortarticle}}
  {\ifthenelse{\boolean{singlecolumn}}{}{
    {\raggedright\baselineskip= 24pt\titlefont \@title\par}%
    \vskip10pt
    {\raggedright \@author\par}
    \vskip8pt
    {\raggedright \addinfofont \@ifundefined{@addinfo}{}{\@addinfo}\par}
    \vskip12pt%
    }}
  {
    \vskip10pt%
    {\raggedright\baselineskip= 24pt\titlefont \@title\par}%
    \vskip10pt
    {\raggedright \@author\par}
    \vskip8pt
    {\raggedright \addinfofont \@ifundefined{@addinfo}{}{\@addinfo}\par}
    \vskip12pt
    {%
    \abscontent
    }%
    \vskip25pt%
  }%
}%
}

\usepackage[flushmargin,ragged]{footmisc}
\renewcommand{\footnoterule}{
  \kern -3pt
  {\color{black50} \hrule width 72pt height 0.25pt}
  \kern 2.5pt
}

\titleclass{\acknow@section}{straight}[\part]
\newcounter{acknow@section}
\providecommand*{\toclevel@acknow@section}{0}
\titleformat{\acknow@section}[runin]
   {\sffamily\normalsize\bfseries\color{macblue}}
   {}
   {0em}
   {#1.}
   []
\titlespacing{\acknow@section}
	{0pt}
	{3.25ex plus 1ex minus .2ex}
	{1.5ex plus .2ex}

\newcommand{\showacknow}{
\@ifundefined{@acknow}{}{
\vskip 3.25ex plus 1ex minus .2ex
\noindent{\sffamily\normalsize\bfseries Acknowledgments.\hspace{1.5ex plus .2ex}}
\small\@acknow}
}

\titleclass{\matmethods@section}{straight}[\part]
\newcounter{matmethods@section}
\providecommand*{\toclevel@matmethods@section}{0}
\titleformat{\matmethods@section}
   {\sffamily\normalsize\bfseries}
   {}
   {0em}
   {#1}
   []
\titlespacing{\matmethods@section}
	{0pt}
	{3.25ex plus 1ex minus .2ex}
	{1.5ex plus .2ex}
\newcommand{\showmatmethods}{
\@ifundefined{@matmethods}{}{\matmethods@section{Materials and Methods}{\small\noindent\@matmethods}}
}

\usepackage{enumitem} 

\makeatletter
\DeclareRobustCommand\code{\bgroup\@noligs\@codex}
\def\@codex#1{\texorpdfstring%
{{\normalfont\ttfamily\hyphenchar\font=-1 #1}}%
{#1}\egroup}
\makeatother

\usepackage{fancyvrb}

\DefineVerbatimEnvironment{Highlighting}{Verbatim}{commandchars=\\\{\}}

\definecolor{codeboxbg}{RGB}{246,246,246}
\global\mdfdefinestyle{codebox}{%
  linewidth=0pt,
  backgroundcolor=codeboxbg,
  innerleftmargin=2pt,
  innerrightmargin=2pt}

\global\mdfdefinestyle{resultbox}{%
  linewidth=0pt,
  backgroundcolor=codeboxbg,
  innerleftmargin=2pt,
  innerrightmargin=2pt}

\setboolean{shortarticle}{true}

\usepackage{float}
\floatstyle{plain}
\newfloat{sigstatement}{b!}{sst}

\additionalelement{%
\afterpage{\begin{sigstatement}
\sffamily
\mdfdefinestyle{pnassigstyle}{linewidth=0.7pt,backgroundcolor=pnasblueback,linecolor=pnasbluetext,fontcolor=pnasbluetext,innertopmargin=6pt,innerrightmargin=6pt,innerbottommargin=6pt,innerleftmargin=6pt}
\@ifundefined{@significancestatement}{}{%
	\begin{mdframed}[style=pnassigstyle]%
	\section*{Significance Statement}%
	\@significancestatement
	\end{mdframed}}
\scriptsize
\@ifundefined{@authorcontributions}{}{\@authorcontributions}
\vskip5pt%
\@ifundefined{@authordeclaration}{}{\@authordeclaration}
\vskip5pt%
\@ifundefined{@equalauthors}{}{\@equalauthors}
\vskip5pt%
\@ifundefined{@correspondingauthor}{}{\@correspondingauthor}
\end{sigstatement}}
}

\usepackage[bitstream-charter]{mathdesign} 

\usepackage{upquote} 

\usepackage{longtable}
\footercontents{\textsf{\bfseries ar\textcolor[HTML]{B31B1B}{X}iv}}

\graphicspath{{Figures/}{./}}

\newcommand{\dif}[1]{\ensuremath{\operatorname{d}\!{#1}}}
\newcommand{\T}{^\top}
\newcommand{\X}{\boldsymbol{X}}
\newcommand{\x}{\boldsymbol{x}}
\newcommand{\Z}{\boldsymbol{Z}}
\newcommand{\z}{\boldsymbol{z}}
\newcommand{\mub}{\boldsymbol{\mu}}
\newcommand{\Sigmab}{\boldsymbol{\Sigma}}
\newcommand{\thetab}{\boldsymbol{\theta}}
\newcommand{\Psib}{\boldsymbol{\Psi}}
\newcommand{\etab}{\boldsymbol{\eta}}

\newcommand{\Gammab}{\boldsymbol{\Gamma}}
\newcommand{\Deltab}{\boldsymbol{\Delta}}
\newcommand{\U}{\boldsymbol{U}}
\newcommand{\B}{\boldsymbol{B}}
\newcommand{\Space}{\mathcal{S}}
\DeclareMathOperator{\Real}{\mathbb{R}}
\DeclareMathOperator*{\argmax}{arg\max}
\DeclareMathOperator{\vech}{vech}

\title{Modal clustering on PPGMMGA projection subspace}

\author{
  \vspace*{2ex}
    \small\textbf{Luca Scrucca}
   \\
  \footnotesize
    Department of Economics \\
    Università degli Studi di Perugia \\
  Via A. Pascoli 20, 06123 Perugia, Italy \\
  \textcolor[HTML]{005392}{\faEnvelope}\;\;\href{mailto:luca.scrucca@unipg.it}{\nolinkurl{luca.scrucca@unipg.it}} \\
  \textcolor[HTML]{A6CE39}{\faOrcid}\;\;\url{https://orcid.org/0000-0003-3826-0484}\\
  }

\begin{abstract}
PPGMMGA is a Projection Pursuit (PP) algorithm aimed at detecting and visualizing clustering structures in multivariate data. 
The algorithm uses the negentropy as PP index obtained by fitting Gaussian Mixture Models (GMMs) for density estimation and, then, exploits Genetic Algorithms (GAs) for its optimization. 
Since the PPGMMGA algorithm is a dimension reduction technique specifically introduced for visualization purposes, cluster memberships are not explicitly provided.
In this paper a modal clustering approach is proposed for estimating clusters of projected data points. In particular, a modal EM algorithm is employed to estimate the modes corresponding to the local maxima in the projection subspace of the underlying density estimated using parsimonious GMMs. Data points are then clustered according to the domain of attraction of the identified modes. Simulated and real data are discussed to illustrate the proposed method and evaluate the clustering performance.
\end{abstract}

\keywords{Gaussian mixtures models; projection pursuit; negentropy index; genetic algorithms; modal EM; modal clustering.}

\begin{document}

\maketitle
\thispagestyle{firststyle}
\ifthenelse{\boolean{shortarticle}}{\ifthenelse{\boolean{singlecolumn}}{\abscontentformatted}{\abscontent}}{}

\section{Introduction}
\label{sec:intro}

Finite mixture models are an important tool in statistical data analysis. Because of their flexibility, they have been successfully used for different purposes, from modelling of the heterogeneity to clustering, from supervised classification to semiparametric density estimation \citep{McLachlan:Peel:2000, McLachlan:etal:2019}.
Model-based clustering is often used as a synonym to indicate the use of mixture models to investigate the presence of clusters in the data.

Projection Pursuit (PP) refers to a collection of algorithms aimed at finding interesting structures in the data \citep{Friedman:Tukey:1974, Friedman:1987}.
Often, the main interest consists in detecting the presence of homogeneous and well separated groups of data.
Recently, a projection pursuit algorithm based on Gaussian Mixture Models (GMMs) has been proposed by \citet{Scrucca:Serafini:2019}. This PP algorithm employs the negentropy obtained from a multivariate density estimated by GMMs as the PP index to be maximized. Optimization of the negentropy index is performed using Genetic Algorithms (GAs).
The resulting PPGMMGA algorithm estimates the optimal orthogonal projection subspace which is able to show the directions of maximal non normality, and for this reason it can be effectively employed to detect and visualize the presence of clustering structures in multivariate data. However, being PPGMMGA a dimension reduction method, no clustering memberships are assigned to the projected data points.

The main goal of this contribution is to propose a clustering approach based on the modal EM algorithm which allows to estimate the modes corresponding to the local maxima of the underlying density of projected data. Thus, groups are  obtained according to the mode to which each data point converges, thereby providing a clustering procedure for the PPGMMGA algorithm.
The modal clustering procedure discussed in this paper follows the definition of clusters proposed by \citet[p. 205]{Hartigan:1975}, according to which "clusters may be thought of as regions of high density separated from other such regions by regions of low density". 

Section~\ref{sec:review} provides a brief review of parsimonious GMMs and of the model-based projection pursuit algorithm used to maximize the negentropy of a GMM. 
Section~\ref{sec:ppgmmmem} contains the modal clustering proposal for applying the modal EM algorithm on the estimated projection subspace. Section~\ref{sec:examples} describes the empirical results obtained from the application of the proposed modal EM algorithm to examples using both synthetic and real datasets. The final section provides some concluding remarks.

\section{Background}
\label{sec:review}

\subsection{Parsimonious Gaussian mixtures}
\label{sec:gmm}

Let $\x$ be a random vector in $\mathbb{R}^p$ with density $f(\x)$, which we assume it can be described by a finite mixture of Gaussian distributions:
\begin{equation}
f(\x ; \thetab) = \sum^{G}_{g=1} \pi_{g} \phi( \x ; \mub_g, \Sigmab_g),
\label{eq:gmm}
\end{equation}
where $\thetab = \{\pi_1, \pi_2, \ldots, \pi_{G-1}, \mub_1, \ldots, \mub_G, \vech(\Sigmab_1), \ldots, \vech(\Sigmab_G) \}$ are the parameters of the mixture model with $G$ components, ($\pi_1, \pi_2, \ldots, \pi_G$) the mixing weights, so that $\pi_g > 0$ and $\sum_{g=1}^{G} \pi_g = 1$, and $\phi(\x ;  \mub_g, \Sigmab_g)$ the underlying multivariate Gaussian density function of the $g$th component having mean $\mub_g$ and covariance matrix $\Sigmab_g$. Note that $\vech(\cdot)$ is the vector operator, which forms a vector by extracting unique elements of a symmetric matrix.

Parsimonious parametrization of covariance matrices for GMMs is available through the eigen-decomposition $\Sigmab_g = \lambda_g \U_g \Deltab_g \U\T_g$, where $\lambda_g = |\Sigmab_g|^{1/p}$ is a scalar which controls the volume, $\Deltab_g$ is the diagonal matrix of normalized eigenvalues of $\Sigmab_g$, such that $|\Deltab_g| = 1$, which controls the shape, and $\U_g$ is the orthogonal matrix of eigenvectors of $\Sigmab_g$ controlling the orientation of the ellipsoid \citep{Banfield:Raftery:1993, Celeux:Govaert:1995}.
Maximum likelihood estimates of the mixture parameters can be obtained using the Expectation-Maximisation (EM) algorithm \citep{Dempster:Laird:Rubin:1977, McLachlan:Krishnan:2008}, whereas information criteria, such as the Bayesian Information Criterion \citep[BIC;][]{Schwartz:1978}, are typically used for model selection. The latter concerns the decision on both the number of mixture components and on which parsimonious parametrization of covariances to adopt. 
Other approaches are also available in the literature to reduce the number of unknown parameters. For instance, mixture of factor analyzers, proposed in \citet{McLachlan:Peel:2000b} and based on the work of \citet{Ghahramani:Hinton:1997}, could be considered here by allowing each Gaussian component in the mixture to be represented in a different lower-dimensional manifold.

\subsection{Model-based projection pursuit}
\label{sec:ppgmmga}

Consider the orthogonal linear projection $\z = \B\T \x \in \mathbb{R}^{d}$, with $d < p$, and the basis $\B \in \mathbb{R}^{p \times d}$, a matrix spanning the dimension reduction subspace $\Space(\B)$.
A remarkable result is the \emph{linear transformation property of GMMs} \citep[see Appendix A in][]{Scrucca:Serafini:2019}, which shows that if $\x$ follows a Gaussian mixture distribution as in \eqref{eq:gmm}, then any linear projection also follows a Gaussian mixture  distribution, so we can write
\begin{equation}
f(\z) = \sum^{G}_{g=1} \pi_{g}\phi( \z ; \B\T \mub_g, \B\T \Sigmab_g \B).
\label{eq:ppgmm}
\end{equation}
Thus, the distribution of the projected data can still be described by a GMM with a suitable transformation of the parameters of the Gaussian components.
The basis $\B$ of the subspace can be estimated by maximizing the \emph{negentropy}, a PP index that summarizes deviations from the least interesting case represented by the Gaussian distribution. The negentropy is defined as
$$
J(\z) = h\big(\phi(\mub_{z},\Sigmab_{z})\big) - h(\z)
       = \frac{1}{2} \log \big( (2\pi e )^d |\Sigmab_{z}| \big)  + \int \log \big( f(\z) \big) f(\z) \dif\z,
$$
where $h(\phi\big(\mub_{z},\Sigmab_{z})\big)$ is the entropy of the marginal multivariate Gaussian distribution with mean $\mub_{z}$ and covariance $\Sigmab_{z}$, and $h(\z)$ is the entropy of the mixture density of the projected data. Efficient approximations for $h(\z)$ are available, such as the \emph{unscented transformation} (UT), so the PP index can be maximized over the set of all orthonormal bases $\B$ using genetic algorithms. For further details see \citet{Scrucca:Serafini:2019}.
This briefly described algorithm is named PPGMMGA, and it is implemented in the R package \texttt{ppgmmga} \citep{Rpkg:ppgmmga}. 

\section{Modal clustering on PPGMMGA projection subspace}
\label{sec:ppgmmmem}

In the modal approach to clustering, clusters are defined as the local maxima of the underlying probability density function. For a recent review of different approaches in modal clustering see \citet{Menardi:2016}, while for a more general discussion on the recent renewed interest on modal procedures see \citet{Chacon:2020}. 
The Modal EM (MEM) algorithm is an iterative algorithm aimed at identifying the local maxima of a finite mixture density function \citep{Li:Ray:Lindsay:2007}. Once the modes have been estimated, clustering of data points is obtained according to the mode to which they converge.
Recently, \citet{Scrucca:2021} generalized the MEM algorithm to efficiently deal with Gaussian mixtures having any parsimonious covariance matrices of the type described in Section~\ref{sec:ppgmmga}. 

To effectively apply the MEM approach on the PPGMMGA projection subspace a density estimate is needed. 
In principle, the mixture density in equation~\eqref{eq:ppgmm} could be used as the underlying density to base the MEM procedure. However, if this is reasonable at the population level, in practice only an estimate of the density \eqref{eq:ppgmm} from sample data is available. Even more important, this estimate is typically chosen among a set of plausible models according to a model selection criterion, such as the BIC \citep{Fraley:Raftery:1998}. 
Since the latter penalizes model complexity, more parsimonious models tend to be selected when the sample size of available data is limited compared to the dimensionality of the problem. However, the adopted parameterization may prove to be inadequate to represent the density of the data on the projection subspace. 

A limiting case of this situation is presented in Section~\ref{sec:sim1}, where a simulated dataset of 100 observations on a 50-dimensional feature space is considered. In this case the Gaussian mixture density selected by BIC has only one component, and if this poses no problem for estimating the PPGMMGA subspace, it clearly fails if used for estimating the modes of the underlying density on the two-dimensional projection subspace. 
For this reason, a different route is needed and, again, a parsimonious Gaussian mixture is called to the rescue. 

\subsection{MEM algorithm}
\label{sec:ppgmmmemalgo}

Consider the PPGMMGA basis $\B$ estimated for a fixed dimensionality $d$. The projected data can be computed as $\z_i = \B\T\x_i$ for all $i=1,\ldots,n$, or equivalently in matrix form as $\Z = \X\B$. The Gaussian mixture density on the projection subspace can be written as
\begin{equation}
f(\z ; \Psib) = \sum^{K}_{k=1} \tau_k \phi( \z ; \etab_k, \Gammab_k),
\label{eq:gmmz}
\end{equation}
where $\Psib = \{\tau_1, \tau_2, \ldots, \tau_{K-1}, \etab_1, \ldots, \etab_K, \vech(\Gammab_1), \ldots, \vech(\Gammab_K) \}$ are the parameters of the mixture model with $K$ components, ($\tau_1, \tau_2, \ldots, \tau_K$) the mixing weights, so that $\tau_k > 0$ and $\sum_{k=1}^{K} \tau_k = 1$, and $\phi(\z ;  \etab_k, \Gammab_k)$ the multivariate Gaussian density function of the $k$th component having mean $\etab_k$ and covariance matrix $\Gammab_k$. 
Following the details provided in Section~\ref{sec:gmm}, the best fitting parsimonious mixture density \eqref{eq:gmmz} can be selected by BIC with respect to both the number of mixture components $K$ and the parsimonious covariance matrices 
$\{\Gammab_k\}_{k=1}^K$. By using the estimated density it is possible to identify the modes in the dimension reduction subspace $\Space(\B)$, thus enabling a modal clustering approach for the identification of homogeneous and separated groups. The MEM algorithm is briefly described below, while further details for interested readers, including a discussion on the number of modes in Gaussian mixtures, are provided in \citet{Scrucca:2021}.

The MEM algorithm starts at initial data points $\z_i^{(0)} = \z_i$ for $i=1,\ldots,n$, then at each iteration $t$, assuming the mixture parameters $\{\tau_k, \etab_k, \Gammab_k\}_{k=1}^K$ as known and fixed, performs the following steps:
\begin{description}

\item{\it E-step} -- update the posterior conditional probability of the projected data point $\z_i$ to belong to the $k$th mixture component using
\begin{equation*} 
\zeta_{ik}^{(t)} = \frac{\tau_k \phi(\z_i^{(t-1)} ;\; \etab_k, \Gammab_k)}{ \sum_{j=1}^K \tau_j \phi(\z_i^{(t-1)} ;\; \etab_j, \Gammab_j) }
\quad\text{for all } k = 1, \ldots, K.
\end{equation*}

\item{\it M-step} -- update the current value of $\z_i$ by solving the optimization problem
\begin{equation*}
\z_i^{(t)} = \argmax_{\z_i \in \Real^d} \sum_{k=1}^K \zeta_{ik}^{(t)} \log \phi(\z_i^{(t-1)} ;\; \etab_k, \Gammab_k).
\end{equation*}
\end{description}
The procedure above is iterated until a convergence criterion is satisfied, or a pre-specified maximum number of iterations is reached.

Note that the objective function in the M-step can be written as
\begin{equation*}
Q(\z_i) = \sum_{k=1}^K \zeta_{ik} \log \phi(\z_i ;\; \etab_k, \Gammab_k),
\end{equation*}
so the gradient and Hessian of this function with respect to the projected vector $\z_i$ are, respectively,
\begin{align*}
\nabla Q(\z_i) & = - \sum_{k=1}^K \zeta_{ik} \Gammab_k^{-1} (\z_i - \etab_k),
\quad\text{and}\\
\nabla^2 Q(\z_i) & = - \sum_{k=1}^K \zeta_{ik} \Gammab_k^{-1}.
\end{align*}
Therefore, the M-step presents a closed-form solution given by
\begin{equation}
\z_i^* = \left( \sum_{k=1}^K \zeta_{ik} \Gammab_k^{-1} \right)^{-1} \sum_{k=1}^K \zeta_{ik} \Gammab_k^{-1} \etab_k.
\label{eq:propsol}
\end{equation}

A direct application of the update equation above may yield large jumps during the initial iterations of the algorithm for data points lying in low-density regions, thus escaping from the region of the attracting mode to the domain of attraction of a different mode. 
In other word, a data point located in a low-density region might see large jumps at the beginning of the iterative algorithm, hence converging to a mode further from its domain of attraction.
For this reason, \citet[Section 3.1]{Scrucca:2021} introduced a step size to get more reliable update at iteration $t$ using a convex linear combination of the solution at previous iteration, $\z_i^{(t-1)}$, and the new proposal $\z_i^*$ from \eqref{eq:propsol}, as follows
\begin{equation}
\z_i^{(t)} = (1-\omega)\, \z_i^{(t-1)} + \omega\, \z_i^*,
\label{eq:sol}
\end{equation}
where $\omega = 1 - \exp\{-0.1 t\}$ is a parameter that controls the step size. Following this approach, at earlier iterations the step size is small, hence updates are done in small steps, but as the number of iterations increase the step size converges to one and the updates become essentially equal to the proposal $\z_i^*$.
The proposed modification does not break the monotonicity property of EM (see Appendix for the proof), although it slows the convergence somewhat by increasing the number of iterations required.

Finally, we note that the execution time can be greatly reduced by batch updating all the data points at once instead of working one point at time, as detailed in \citet{Scrucca:2021}.

\subsection{Some final comments}
\label{sec:ppgmmmemcom}

In the model-based clustering approach, clusters are defined by the underlying distribution used for the mixture components. For instance, GMMs assume ellipsoidal group structures. Then, the so-called maximum a posteriori (MAP) principle is used for clustering.
On the contrary, as discussed in Section~\ref{sec:intro}, modal clustering is an attempt at addressing the definition of clusters proposed by \citet{Hartigan:1975}. 
If data points projected onto the PPGMMGA projection subspace show ellipsoidal structures and are well-separated then the model-based clustering approach using GMMs and the modal-clustering via MEM algorithm produce equivalent results. 
However, in case the groups show non-Gaussian structures or substantially overlap on the projection subspace, the final clustering might be different. This mostly reflects the different definition adopted by the two approaches regarding what a cluster is. An example of such a case is discussed in Section~\ref{sec:coffee}.

\section{Data analysis examples}
\label{sec:examples}

In this section we apply the proposed methodology to simulated and real datasets.

\subsection{Simulated data with clustering on a subset of attributes}
\label{sec:sim1}

We consider a reduced version of the two-group synthetic dataset discussed in \citet{Friedman:Meulman:2004}. For the first group 85 observations are randomly generated from a 50-dimensional standard Gaussian distribution. The second group of 15 observations are also drawn from a 50-dimensional Gaussian distribution, but its first $p_0=15$ attributes have mean $1.5$ and standard deviation $0.2$, whereas the remaining $50-p_0$ attributes have zero mean and unit standard deviation. Thus, the two groups differ only on the first $p_0$ attributes. 

\begin{figure}[htb]
\includegraphics[width=\textwidth]{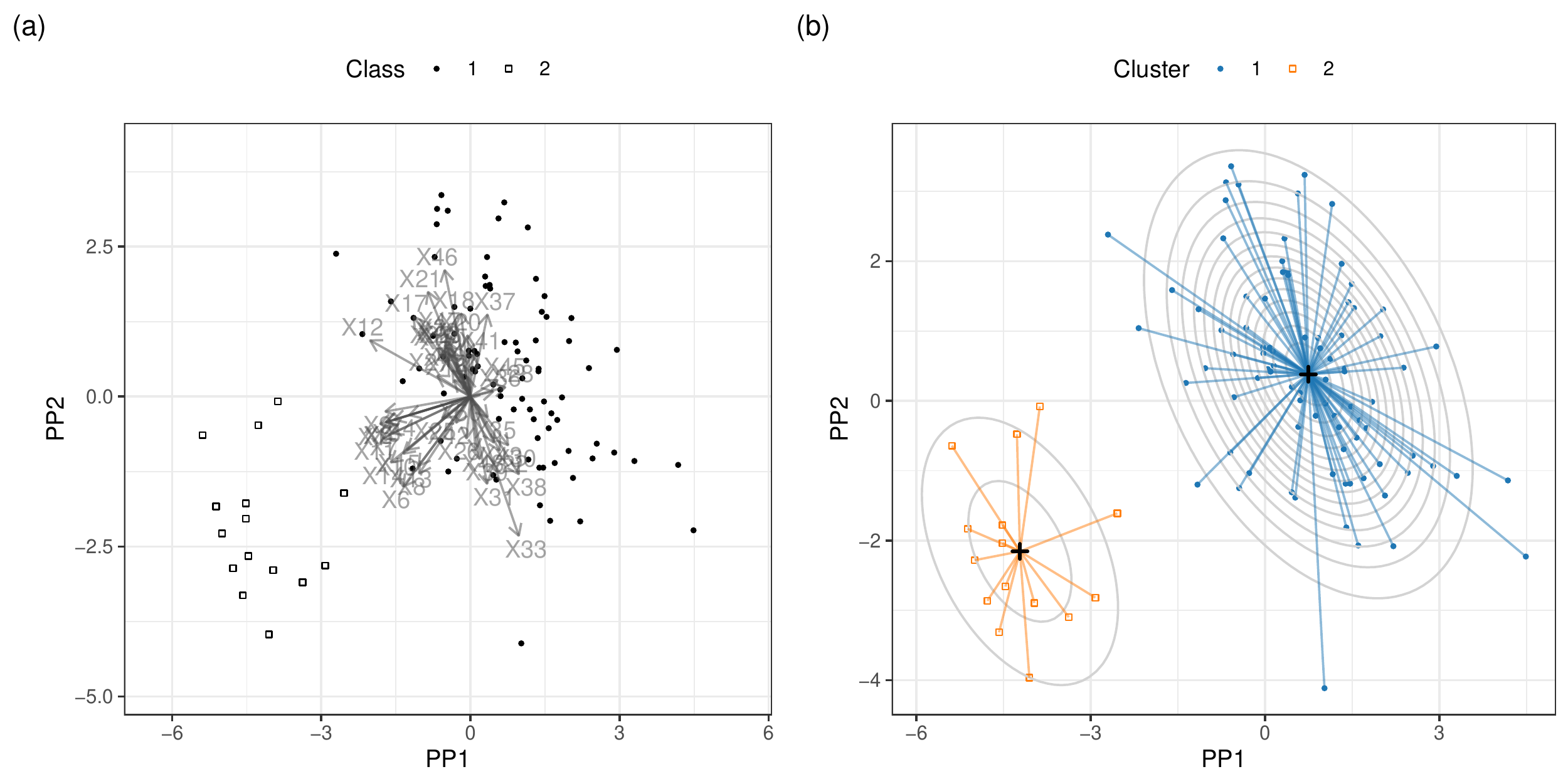}
\caption{Plot of two-group synthetic data points projected along the PPGMMGA directions with points marked according to the known class in panel (a), and using the modal clustering solution on panel (b). The last graph also displays the MEM iteration paths toward the attracting modes represented using the $+$ symbol.}
\label{fig:sim1}
\end{figure}

Figure~\ref{fig:sim1}a shows the data projected onto the 2-dimensional projection subspace estimated by the PPGMMGA algorithm, which clearly reveals the two-group clustering structure. 
It is interesting to note that the GMM selected by BIC on the original 50-dimensional features space has a single Gaussian component. 
In this case, an attempt at clustering on the original features space would produce a solution consisting of a single cluster.
Nonetheless, the estimated directions that maximize the negentropy are able to show the underlying structure as shown in Figure~\ref{fig:sim1}a. 

Iteration paths of MEM algorithm for each data point toward the attracting mode are depicted in Figure~\ref{fig:sim1}b. The resulting modal clustering solution obtained on the projection subspace is perfectly able to recover the two groups of observations. 

\subsection{Rectangular Gaussian block clusters in three dimensions with noise attributes}
\label{sec:sim2}

In this second example, we consider a simulated dataset made up of 400 observations on 8 variables. The first three variables are generated from an equal weight mixture of eight Gaussian distributions with mean vectors at the corners of a rectangular block in three dimensions. The remaining five variables only contain some Gaussian noise. Details about the mean vector for each mixture component and the common covariance matrix are provided in \citet[Appendix A]{Bolton:Krzanowski:2003}.

\begin{figure}[htb]
\centering
\includegraphics[width=\textwidth]{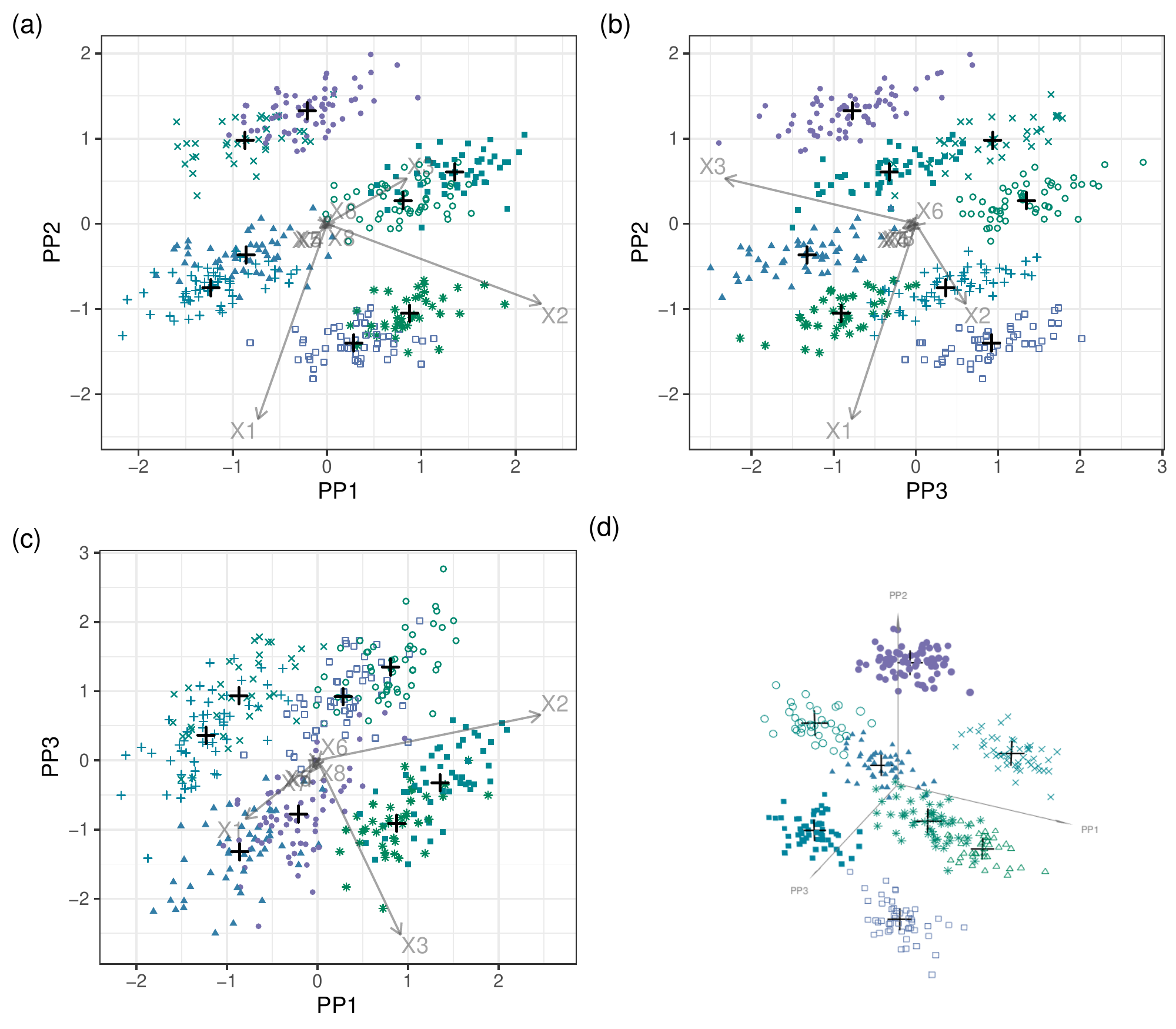}
\caption{Pairwise scatterplots (panels (a) to (c)) and a static view of 3D scatterplot (panel (d)) of PPGMMGA projections with data points marked according to the clusters assigned by the MEM procedure and the corresponding modes, represented using $+$ symbol.}
\label{fig1:simdata}
\end{figure}

We consider the PPGMMGA estimation of the projection subspace for varying dimensions. The negentropy values for each solution with $d$ from 1 up to 5 suggest a 3-dimensional projection subspace as a reasonable choice (see Table~\ref{tab1:simdata}). Figure~\ref{fig1:simdata} shows some views of the data points projected along the estimated PPGMMGA directions with modes and data points marked according to the cluster memberships estimated by the MEM algorithm.
The 2-dimensional scatterplots in panels (a)--(c) show pairwise projections for the estimated PPGMMGA directions, with arrows indicating the biplot vectors corresponding to the original features. As it can be seen, only the first three variables contribute to the definition of the PPGMMGA directions (see also the coefficients of the estimated projection basis reported in Table~\ref{tab2:simdata}).
Finally, Figure~\ref{fig1:simdata}d shows a static view of the data points projected onto the estimated 3-dimensional projection subspace. 

Table~\ref{tab1:simdata} contains, in addition to the negentropy for increasing dimensionality of the projection subspace, the number of modes found by the MEM algorithm and the accuracy of the proposed modal clustering procedure. This is measured by the Adjusted Rand Index \citep[ARI; ][]{Hubert:Arabie:1985}, a measure of agreement between two partitions, which attains unit value when the two partitions perfectly agree. Overall, the MEM algorithm is able to recover almost perfectly the true clusters for $d \ge 3$.

\begin{table}[htb]
  \centering
  \begin{minipage}[b]{.48\textwidth}
    \centering
    \setlength{\tabcolsep}{2ex}
    \caption{PPGMMGA negentropy, number of modes ($m$), and adjusted Rand index (ARI) for increasing dimension ($d$) of the projection subspace.}
    \label{tab1:simdata}
    \begin{tabular}{rrrr}
    \hline
    $d$ & Negentropy & $m$ & ARI \\ 
    \hline
      1 & 0.5609 &  3 & 0.4244 \\ 
      2 & 1.0832 &  7 & 0.8653 \\ 
      3 & 1.5075 &  8 & 0.9942 \\ 
      4 & 1.5271 &  8 & 0.9942 \\ 
      5 & 1.5351 &  8 & 0.9942 \\ 
    \hline
    \end{tabular}
    \end{minipage}%
    \hfill
    \begin{minipage}[b]{.48\textwidth}
    \centering
    \caption{Estimated PPGMMGA projection basis for the $d=3$ solution.}
    \label{tab2:simdata}
    \begin{tabular}{lrrr}
    \hline
      & PP1 & PP2 & PP3 \\ 
    \hline
    $X_{1}$ & -0.2885 & -0.9041 & -0.3080 \\ 
    $X_{2}$ & 0.8964 & -0.3706 & 0.2394 \\ 
    $X_{3}$ & 0.3343 & 0.2101 & -0.9168 \\ 
    $X_{4}$ & -0.0097 & -0.0203 & -0.0425 \\ 
    $X_{5}$ & -0.0184 & -0.0149 & -0.0361 \\ 
    $X_{6}$ & 0.0099 & 0.0026 & 0.0111 \\ 
    $X_{7}$ & -0.0297 & -0.0203 & -0.0626 \\ 
    $X_{8}$ & 0.0020 & -0.0097 & -0.0109 \\  
    \hline
    \end{tabular}
    \end{minipage}
\end{table}

\clearpage

\subsection{Chironomus larvae data}
\label{sec:chiro}

This dataset comes from a study on morphometric attributes of Chironomus larvae \citep{Atchley:Martin:1971}. A sample of 149 larvae was collected from 3 different species: cloacalis, frommeri and staegeri. For each larva, 17 features of the larval head capsula were measured; for a detailed description of the dataset we refer the reader to \citet{Montanari:Viroli:2010}.

\begin{figure}[htb]
\includegraphics[width=\textwidth]{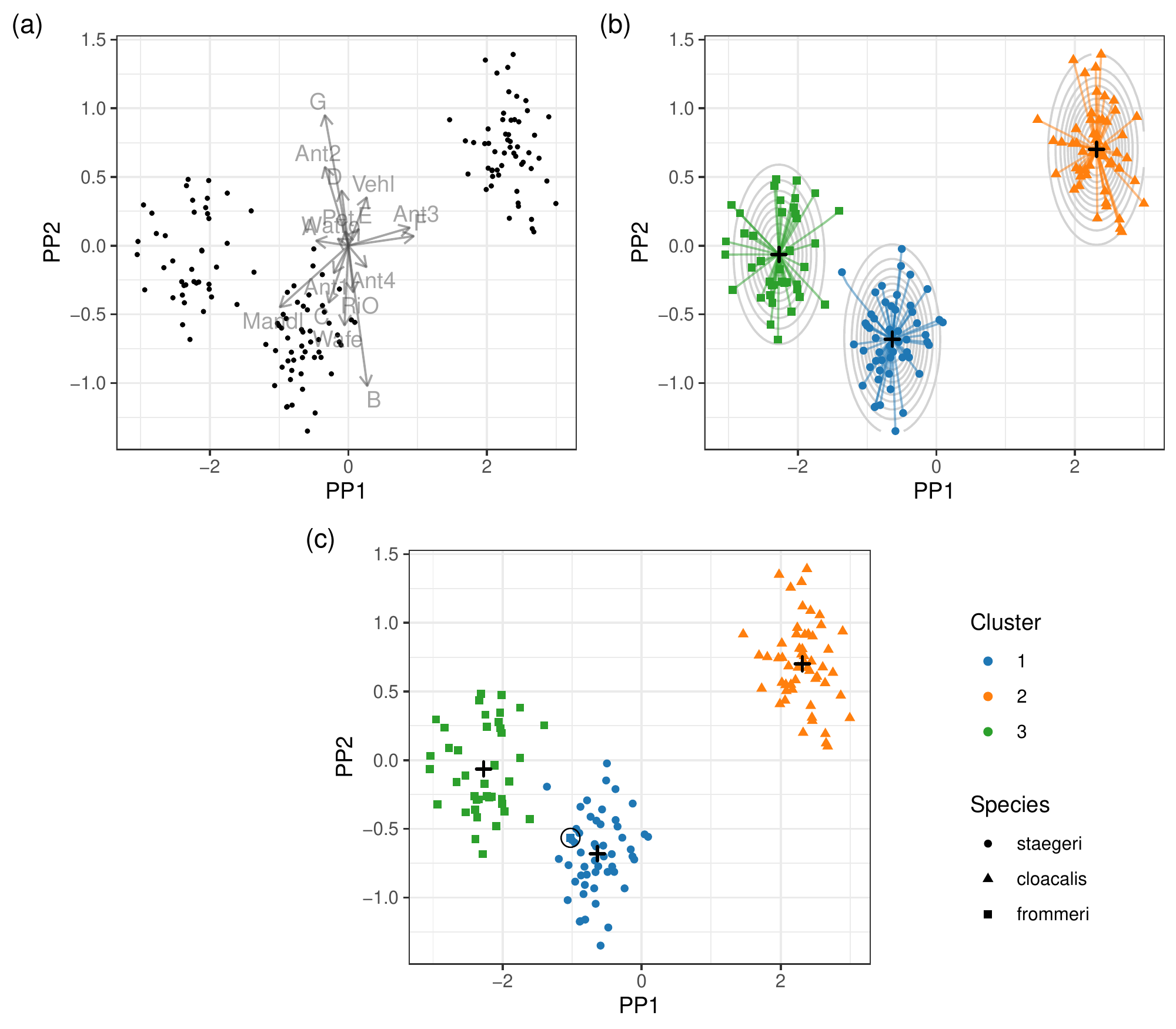}
\caption{Chironomus larvae data graphs: (a) plot of data points projected along the PPGMMGA directions; (b) plot of the modal clustering solution identified by the MEM algorithm with the estimated modes represented by a $+$ symbol; (c) plot of data points marked according to the true species of the larvae with different symbols, and different colours for the identified clusters (circled point represents the misclassified observation).}
\label{fig:chiro}
\end{figure}

Figure~\ref{fig:chiro}a shows the data projected from the original 17-dimensional feature space to the 2-dimensional projection subspace estimated by the PPGMMGA algorithm. The estimated modes and the corresponding modal clustering solutions are shown in Figure~\ref{fig:chiro}b, together with the paths induced by the MEM algorithm. The 3-mode clustering structure identified is then compared with the true larvae species in Figure~\ref{fig:chiro}c, showing a quite good agreement with only one misclassified observation and an adjusted Rand index of $0.981$ (slightly larger than the ARI reported in \citet{Montanari:Viroli:2010}).

\clearpage

\subsection{Coffee data}
\label{sec:coffee}

\begin{figure}[htb]
\centering
\includegraphics[width=\textwidth]{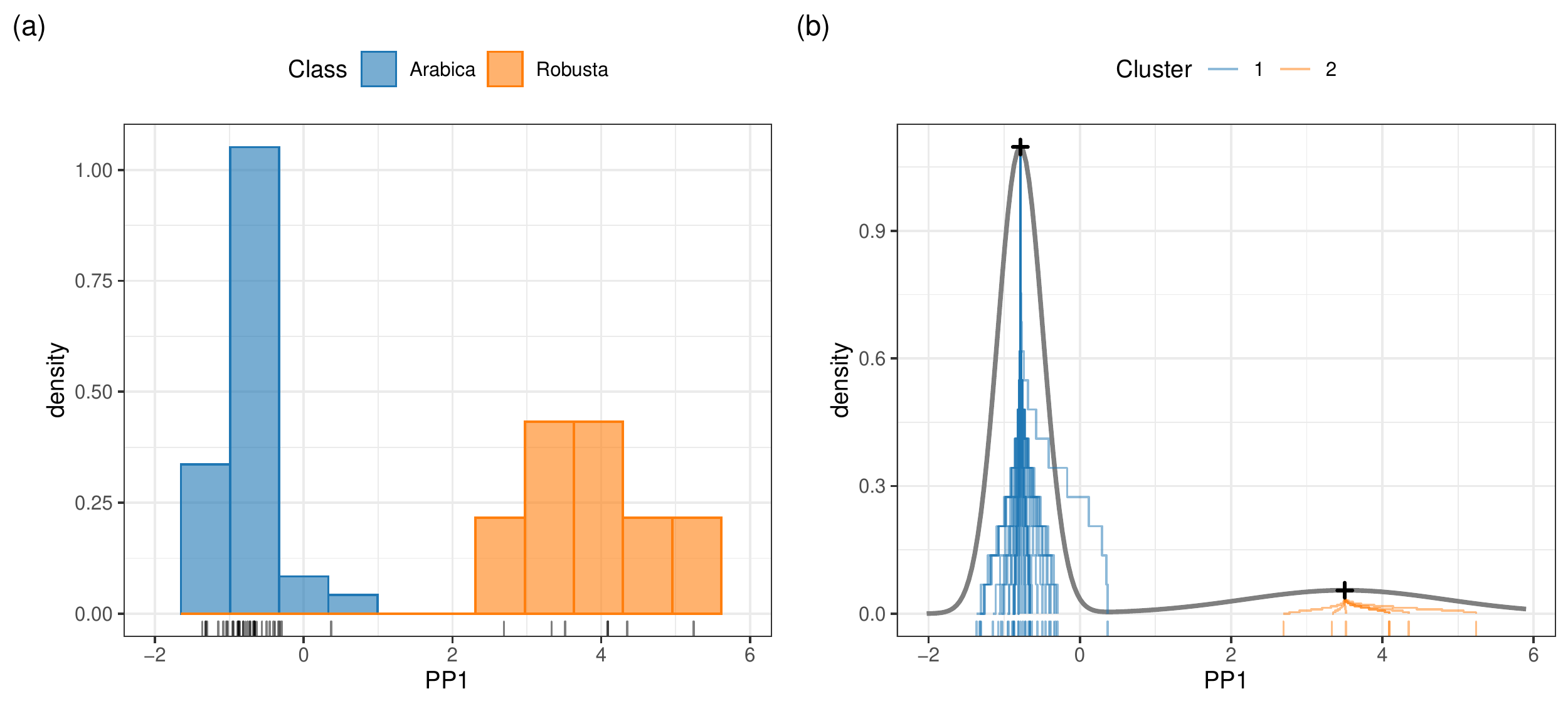}
\caption{Histogram of coffee data points projected along the first PPGMMGA direction marked by the true variety of coffee on panel (a). Underlying density estimated by GMM employed by the MEM algorithm on panel (b), with step lines showing the iteration paths toward the attracting modes represented by a $+$ symbol. Rug points along the $x$-axis on both graphs represent the projected data points.}
\label{fig1:coffee}
\end{figure}

The coffee dataset \citep{Streuli:1973} contains 12 chemical measurements on 43 samples of Arabica or Robusta coffee from 29 countries. \citet{Scrucca:Serafini:2019} analysed the coffee data by fitting a 1-dimensional PPGMMGA model. 
Figure~\ref{fig1:coffee}a shows the data points projected along the estimated direction, clearly suggesting that coffee varieties can be clustered on this one-dimensional subspace.  
Graph on Figure~\ref{fig1:coffee}b reports the underlying density estimated by GMM and used by the MEM algorithm, together with the ascending paths for each projected data point toward the closest mode. The clustering obtained perfectly recovers the latent coffee varieties.

\begin{figure}[htb]
\centering
\includegraphics[width=\textwidth]{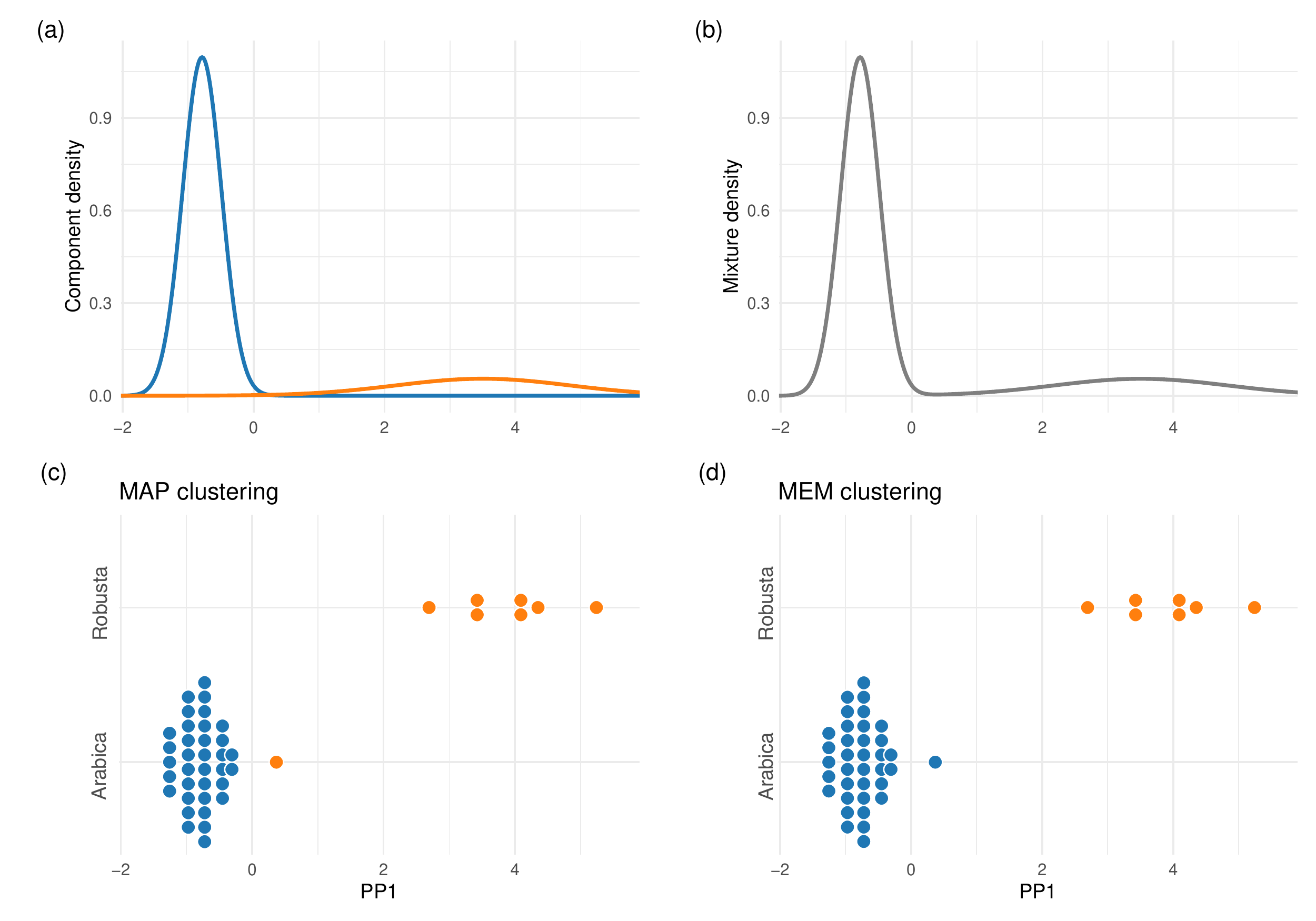}
\caption{%
A graphical comparison of model-based clustering vs modal clustering on the 1-dimensional PPGMMGA projection for the coffee dataset. 
The two Gaussian components of the mixture (weighted by the mixing proportion) are shown in panel (a), whereas the corresponding mixture density is reported in panel (b).
The clustering solutions obtained by applying the MAP principle and the MEM algorithm are shown, respectively, in panels (c) and (d).}
\label{fig2:coffee}
\end{figure}

In Section~\ref{sec:ppgmmmemcom} the modal clustering approach via MEM algorithm and the model-based clustering approach using the MAP principle were discussed and contrasted. 
The considerations made therein can now be illustrated in practice using the coffee dataset.  
In particular, Figure~\ref{fig2:coffee} reports the clustering results obtained using the two approaches on the 1-dimensional PPGMMGA projection.
Panel (a) shows the two Gaussian components of the mixture estimated on the first PPGMMGA direction, with the corresponding MAP clustering on panel (c). On the contrary, panel (b) contains the estimated mixture density, with the corresponding MEM clustering on panel (d). 
According to the MAP principle, data points are assigned to the cluster whose corresponding component density (weighted by the mixing proportion) is the largest. Note that one data point is assigned to the group of Robusta despite being of type Arabica and quite far from the other data points assigned to the same cluster.
However, this appears reasonable under the model-based perspective because the estimated conditional probability for that point is larger for Robusta than for Arabica (and this is due to the large estimated variance for the second component).
By adopting, instead, a modal clustering perspective, the mixture density must be used to identify the modes, and data points are assigned to the ``closest'' clusters represented by such modes. In this case, the classification obtained perfectly matches the true coffee varieties. Indeed, the previously misclassified data point is now assigned to the Arabica cluster because its domain of attraction is the same as for the other Arabica coffees. 

\clearpage

\section{Concluding comments}

PPGMMGA is a linear projection method introduced by \citet{Scrucca:Serafini:2019} with the aim of visualizing multivariate data when a clustering structure is sought. 
The PPGMMGA algorithm estimates the directions of maximal nonnormality as measured by the negentropy PP index. 
Being a linear mapping from a potential high-dimensional features space to a subspace of reduced dimensionality, it involves linear transformations, such as rotation and reflection, of the original data. 
Thus, if a clustering structure is present in the original features space, PPGMMGA should preserve it and, hopefully, magnify it.

However, if visualization is an important step in understanding the underlying structure of the data, an automatic procedure for the identification of homogenous groups would be desirable.
In this paper we introduce a modal clustering approach to address this issue.
Working on the projection subspace estimated by the PPGMMGA method, a modal EM algorithm is proposed for identifying the modes of the density estimated by fitting parsimonious Gaussian mixture models.

The proposed approach can, in principle, be applied to any dimension reduction method, such as local linear embedding \citep[LLE;][]{Roweis:Saul:2000}, t-distributed stochastic neighbor embedding \citep[t-SNE;][]{VanDerMaatenHinton:2008} and uniform manifold approximation and projection \citep[UMAP;][]{McInnes:Healy:Melville:2018}. We defer the investigation of these extensions to future work.

\bigskip\bigskip

\begin{appendix}
\section*{Appendix}
\label{appendix}

We prove here that the inclusion of the step size in equation \eqref{eq:sol} does not break the monotonicity property of MEM, i.e. $f(\z_i^{(t)}) \ge f(\z_i^{(t-1)})$ for any iteration $t$.
First, we note that by the ascending property of the MEM algorithm for any mixture we have that $f(\z_i^{*}) \ge f(\z_i^{(t-1)})$ \citep[see][Appendix A]{Li:Ray:Lindsay:2007}, where $\z_i^{*}$ is the proposal from equation \eqref{eq:propsol} at current iteration $t$, $\z_i^{(t-1)}$ is the solution at iteration $t-1$, and $f()$ is the mixture density \eqref{eq:gmmz}. 
Then, at iteration $t$ of the algorithm we have
\begin{align*}
f( \z_i^{(t)} ) & = f( (1-\omega) \z_i^{(t-1)} + \omega \z_i^{*} ) && \\
& \ge (1-\omega) f(\z_i^{(t-1)}) + \omega f(\z_i^{*}) && \text{(by the monotonocity along the learning path)} \\
& = f(\z_i^{(t-1)}) + \omega (f(\z_i^{*}) - f(\z_i^{(t-1)}) ) && \\
& \ge f(\z_i^{(t-1)}) && \text{(since $f(\z_i^{*}) - f(\z_i^{(t-1)}) \ge 0$ and $\omega \ge 0$)} 
\end{align*}

\end{appendix}

\bigskip\bigskip

\bibliography{ppgmmgaMEM}
\end{document}